\documentclass[twocolumn,showpacs,preprintnumbers,amsmath,amssymb]{revtex4}
\input psfig.sty
\usepackage{graphicx}% Include figure files
\usepackage{dcolumn}% Align table columns on decimal point
\usepackage{bm}% bold math
\usepackage{latexsym}
\usepackage{epsfig}
\begin{document}
%%%%%%%%%%%%%%%%%%%%%%%%%%%%%%%%%%%%%%%%%%%%%%%%%%%%%%%%%%%%%%%
 \newcommand{\bq}{\begin{equation}}
 \newcommand{\eq}{\end{equation}}
 \newcommand{\bqn}{\begin{eqnarray}}
 \newcommand{\eqn}{\end{eqnarray}}
 \newcommand{\nb}{\nonumber}
 \newcommand{\lb}{\label}
%%%%%%%%%%%%%%%%%%%%%%%%%%%%%%%%%%%%%%%%%%%%%%%%%%%%%%%%%%%%%%%%
\title{Black Hole Formation with an Interacting Vacuum Energy Density: Curvature Effects}
%\title{Can black hole formation be avoided by a growing vacuum energy density?}
\author{E. L. D. Perico$^{1}$}\email{elduartep@usp.br} 
\author{J. A. S. Lima$^2$}\email{limajas@astro.iag.usp.br}
\author{M. Campos$^{2,3}$}\email{m.campos@usp.br}

\vskip 0.5cm
\affiliation{$^1$Instituto de F\'{i}sica, Universidade de S\~ao Paulo, 05508-090 S\~ao
Paulo, SP, Brasil,}
\affiliation{$^2$Departamento de Astronomia, Universidade de S\~ao Paulo, 05508-900 S\~ao
Paulo, SP, Brasil}
\affiliation{$^3$Departamento de F\'\i sica, Universidade Federal de Roraima, 69304-000 Boa Vista, RR, Brasil.}
%\affiliation{$^3$Departamento de Astronomia, Universidade de S\~ao Paulo, 05508-900 S\~ao
%Paulo, SP, Brasil}
%\affiliation{$^2$Instituto de F\'{i}sica, Universidade de S\~ao Paulo, S\~ao Paulo, SP, Brasil}

%
%
%\pacs{Black Holes, General Relativty, Decaying Vacuum}
%\keywords{Dark energy, cosmic distance, Inhomogeneity Parameter}
\date{\today }
\begin{abstract}
The  gravitational collapse of a spherically symmetric massive core of a star  in  which the fluid component
is interacting with a growing vacuum energy density filling a FLRW type geometry with an arbitrary curvature parameter
is investigated. 
The complete set of exact solutions for all values of the free parameters are obtained and the  influence  of the curvature term on the collapsing time, black hole mass and other physical quantities are also discussed in detail. We show that for the same initial conditions the total black hole mass depends only on the effective matter  density parameter (including the vacuum component). It is also shown that the analytical condition to form a black hole i.e. the apparent horizon is not altered by the contribution of the curvature terms, however, the remaining physical quantities are quantitatively modified. 
\end{abstract}
\vspace{.7cm}
\maketitle
%
%%%%%%%%%%%%%%%%%%%%%%
\section{Introduction}
%%%%%%%%%%%%%%%%%%%%%%
%
\renewcommand{\theequation}{1.\arabic{equation}}
\setcounter{equation}{0}

In the current view of cosmology, the present accelerating stage of the Universe is caused by the dark energy component,  usually  represented  by a cosmological constant $\Lambda$ \cite{Riess,Komatsu}. Its contribution to the Einstein Field Equations (EFE) is the same of a perfect simple fluid with constant equation of state (EoS) parameter $\omega \equiv p/\rho=-1$, being interpreted as the net energy density stored 
on the vacuum state of all quantum fields pervading the observed Universe \cite{rev1}.

Although in agreement with the existing astronomical
observations (both at background and perturbative levels), $\Lambda$ is plagued with  the longstanding cosmological constant problem, 
i.e. the huge discrepancy ($\sim$ 120 orders of magnitude) between the theoretical expectations
($\rho_{v} \sim 10^{106} eV^4$) from quantum field theory  by assuming that the natural cutoff is related to gravity (Planck's energy) and the present day ($\rho_{v} \sim 10^{-16} eV^4$) cosmological bounds \cite{Zeldovich67,zee85,Weinberg}. The simplest manner to alleviate such a  
problem is to introduce some dynamics, or equivalently, to replace its constant value, say,  $\Lambda_0$, by a 
time dependent quantity, $\Lambda (t) \equiv 8\pi G\rho_v (t)$, where $\rho_v$ is the vacuum energy density 
(from now on a subscript zero denotes the present day value of a quantity). In this way, the incredible small value of $\Lambda$  is the result of a continuous decaying vacuum process i.e. the value of $\Lambda$ is small nowadays because the Universe is too old. 

Many sources for a dynamical vacuum energy density have been proposed in the literature \cite{sources}, and several phenomenological models based on different decay laws for $\Lambda(t)$ were also discussed even before the discovery of the acceleration of the Universe \cite{OT86,L1,CLW,L2,L3}. Decay $\Lambda(t)$ models were the predecessors of all interacting dark  energy models being until nowadays a very  active field of the current research \cite{L5,new1,ML02,EA1,Harko11,LBS2012}.  
%There are also some attempts to represent 
%out of equilibrium dynamical $\Lambda$  models by a scalar field \cite{ML02,EA1}, as well as 
%based on a Lagrangian description \cite{Lag1}.  

As happens with the cosmic history,  a dynamical $\Lambda (t)$-term  can also affect the formation of black holes. However, for each evolving system, the interacting vacuum  behaves in a quite different  manner. In the expanding Universe, for instance,  the vacuum energy density is a continuously time dependent decreasing function  while for  black hole formation $\rho_v (t)$ is a growing quantity in the course of the collapsing process. In the later case, one may ask whether the increasing repulsive gravitational force (due to the negative vacuum pressure) may prevent the ultimate formation of a singularity.

Recently, Campos and Lima \cite{CL12} (henceforth paper I) discussed the formation of black holes (and naked singularities) during  
the gravitational collapse  of a fluid interacting with a time-varying vacuum in the context of a flat Friedmann-Lemaitre-Robertson-Walker (FLRW)  geometry. 
%In their analysis it was assumed that the star medium is a perfect fluid obeying the EoS $p_f = \omega \rho_f$ while the interacting vacuum was defined by the constancy of the %fraction $\beta = \rho_v/(\rho_v + \rho_f)$. 
For given initial conditions, they solved analytically the basic equations describing  the evolution of the two-fluid interacting 
mixture and analyzed  the development of the apparent horizons before the formation of the singularity.  It was shown that 
a time-varying vacuum energy density increases the collapsing time but, in general, 
it cannot prevent the formation of black holes. Their results also suggested that the cosmic censorship hypothesis (CCH), at least in its weak form, can generically  violated in the presence of a time varying vacuum due to the formation of naked singularities. In this concern, many authors have discussed how naked singularities can observationally be distinguished from black holes through strong gravitational lensing effects and the physics of accretion disks \cite{naked}.

In this paper,  we go one step further by analyzing the influence of the curvature on the results derived in Paper I. As we shall see, following a unified method first proposed by Assad and Lima \cite {AL88} for a one component expanding simple fluid, we obtain  the complete set of collapsing exact solutions for all values of the free parameters describing  the interacting mixture.  The influence  of the curvature on the collapsing time, black hole mass and other physical quantities are also discussed in detail. In particular, it is found that for the same initial conditions the black hole mass depends only on the effective matter  density parameter and that naked singularities are also formed for a large interval of the physical parameters. It is also shown that the analytical condition to form a black hole i.e. the apparent horizon is not altered by the contribution of the curvature terms, however, the remaining physical properties are quantitatively modified.   

%Another closely related issue is the possible influence of $\Lambda(t)$  on the cosmic censorship hypothesis (CCH), as well 
%as on the nature of the singularity.  
%In its weak form, this conjecture eliminates the occurrence of naked singularities in the spherical gravitational 
%collapse whereas its strong version states that all singularities in any realistic spacetime are never 
%visible to a distant observer because are hidden behind an event horizon \cite{Penrose}. Since the 
%earlier counter example to the CCH discussed by Papapetrou \cite {Papa}, the emergence of naked  
%singularities or black holes has been intensively investigated in the  
%literature, including the effect of different material components \cite{Collapse}. However, as far as 
%we know, the possible influence of a time varying $\Lambda$-term in the last stages of a collapsing 
%system (including the formation of a trapped surface and naked singularities) has not been analyzed in the literature. In principle, this is an important issue due two 
%combined effects: (i) unlike what happens in an expanding Universe, the energy density of a coupled vacuum component 
%grows in the course of the gravitational contraction, and (ii) since the vacuum pressure is negative  %
%and  generates repulse gravity, potentially, it might alter significantly the late stages of any collapsing matter distribution. 

%
%%%%%%%%%%%%%%%%%%%%%%%%%%%%%%%%%%%%%%%%%%%%%%%%%%%%%%%%%%%%%%%%%%%
\section{Collapsing star with variable - $\Lambda (t)$}
%%%%%%%%%%%%%%%%%%%%%%%%%%%%%%%%%%%%%%%%%%%%%%%%%%%%%%%%%%%%%%%%%%%
%
\renewcommand{\theequation}{2.\arabic{equation}}
\setcounter{equation}{0}
\subsection{Geometry and Composition of the Collapsing Star Medium}

To begin with, we first remark that the basic discussion here is related to black holes and naked singularities formed 
from collapsing star cores. In this way, the formation process involving supermassive black holes like the ones found in the galactic centers will not be investigated here. 

Let us now consider that the massive core of a star medium is formed by a mixture of a isotropic simple  fluid 
plus a growing vacuum energy density. Inside the core it will be assumed that 
the spacetime is described by a generic FLRW geometry ($c=1$): 
\bq
\lb{2.1}
ds^{2}_{-} = dt^{2} - a^{2}\left(t \right)\left(\frac{dr^{2}}{{1-k\,r^2}} + r^{2} d\Omega^{2}\right), 
\eq
where  $a\left(t \right)$ is the scale factor, $k = 0,\pm 1$ is the curvature parameter and $d\Omega^{2} \equiv d\theta^{2} + \sin^{2}\theta d\varphi^{2}$ 
is the area element on the unit sphere. Such an approximation must work at least for the late stages of the collapsing process. The complete spacetime may be divided into 3 different regions, namely: $V^{\pm}$ and $\Sigma$, where $V^{+}$($V^{-}\;$) is  the exterior (interior)  of the massive core, and $\Sigma$ denotes the surface of the massive collapsing core. Following standard lines, in this paper we shall focus our attention mainly in the spacetime  inside the star core \cite{CaiWang,CL12}.  

The EFE inside the star medium can be written as:
\begin{equation} \label{EE}
G_{-}^{\mu \nu} = 8\pi G \left[{T_{-}^{\mu\nu}}_{(f)} + {T_{-}^{\mu\nu}}_{(v)}\right],
\end{equation}
where $G_{-}^{\mu \nu}$ is the Einstein tensor  and  ${T_{-}^{\mu \nu}}_{(f)}$,  ${T_{-}^{\mu \nu}}_{(v)}$ are the energy-momentum
tensor (EMT) of the fluid component and vacuum, respectively.  ${T_{-}^{\mu \nu}}_{(v)} \equiv \rho_{v} g_{-}^{\mu \nu}$,  where $\rho_v = \Lambda(t)/8\pi G$. 

The Einstein tensor is divergenceless, and, therefore,  the EFE imply that a variable-$\Lambda(t)$ is possible if at least one of the two following conditions are satisfied: (i) ${T_{-}^{\mu \nu}}_{(f)}$ is not separately conserved, i.e., ${T_{-}^{\mu \nu}}_{(f)};_{\nu} \neq 0$, and (ii) ${T_{-}^{\mu \nu}}_{(f)};_{\nu} = 0$ but $G$ is a time dependent quantity \cite{VariableG}. In this paper  we assume that the fluid and vacuum components are interacting and G is constant. 

\begin{figure*}[th]\label{fscale}
                 \centerline{\hspace{0.4cm}\psfig{figure=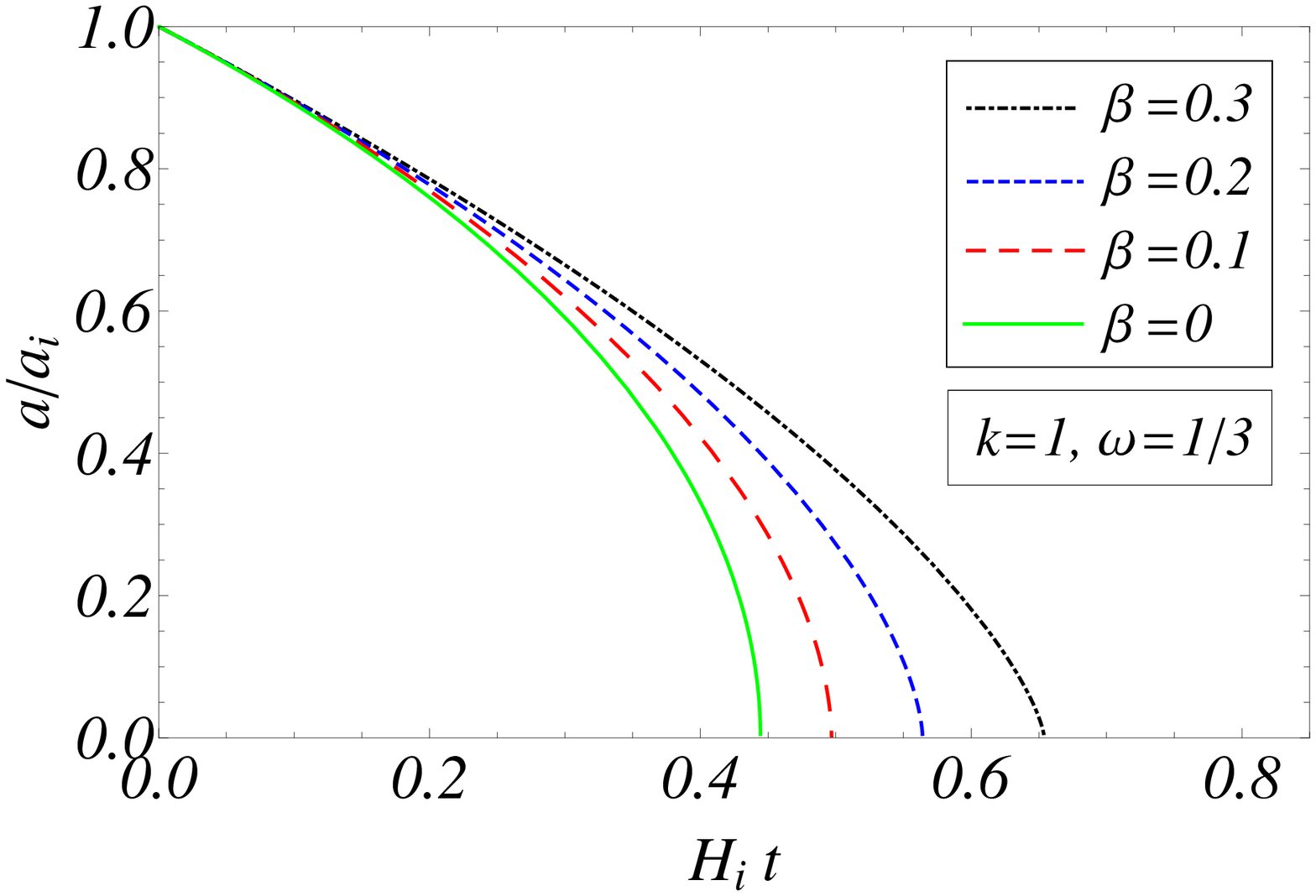,width=2.7truein,height=2.truein}\hspace{-1cm}
								\psfig{figure=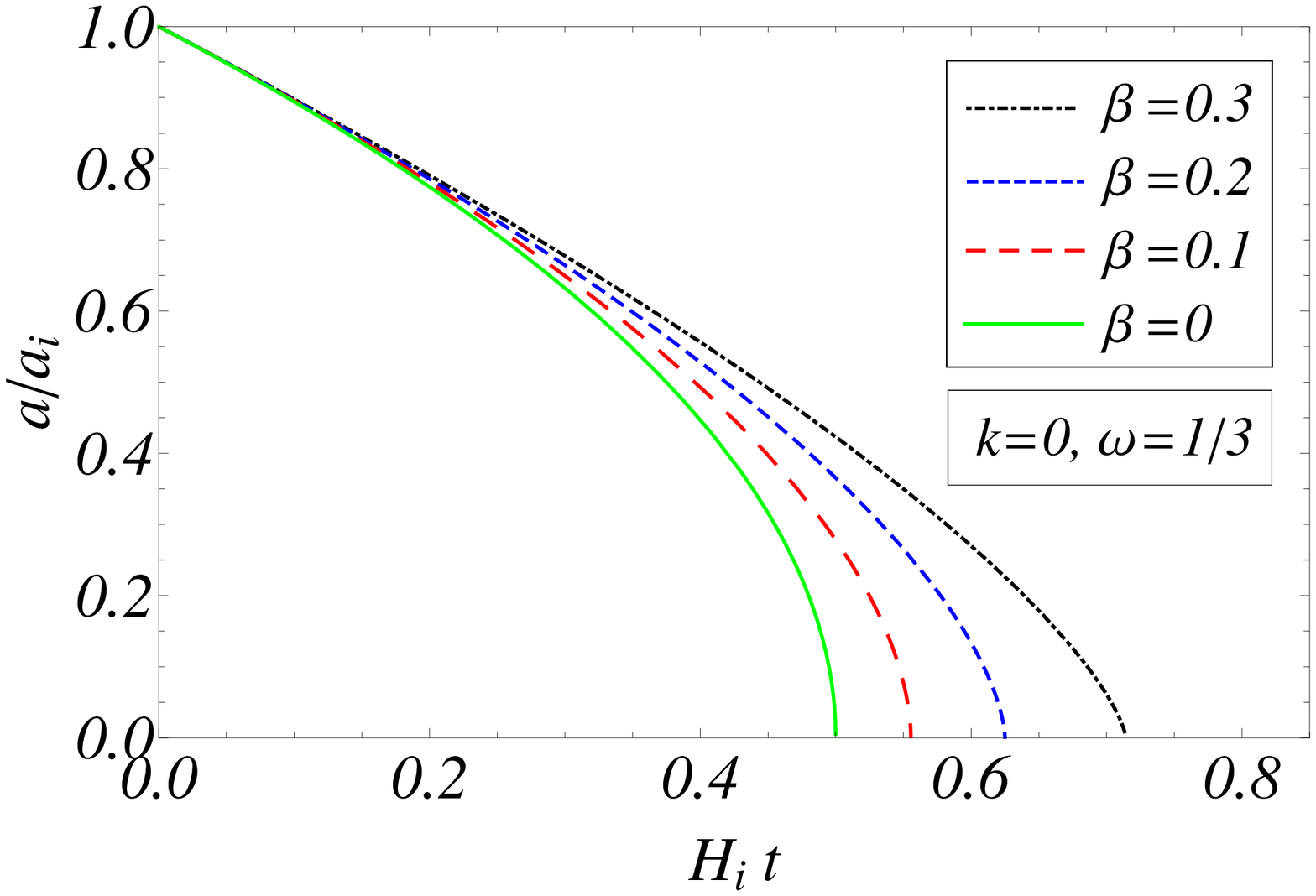,width=2.7truein,height=2.truein}\hspace{-1cm}
                    \psfig{figure=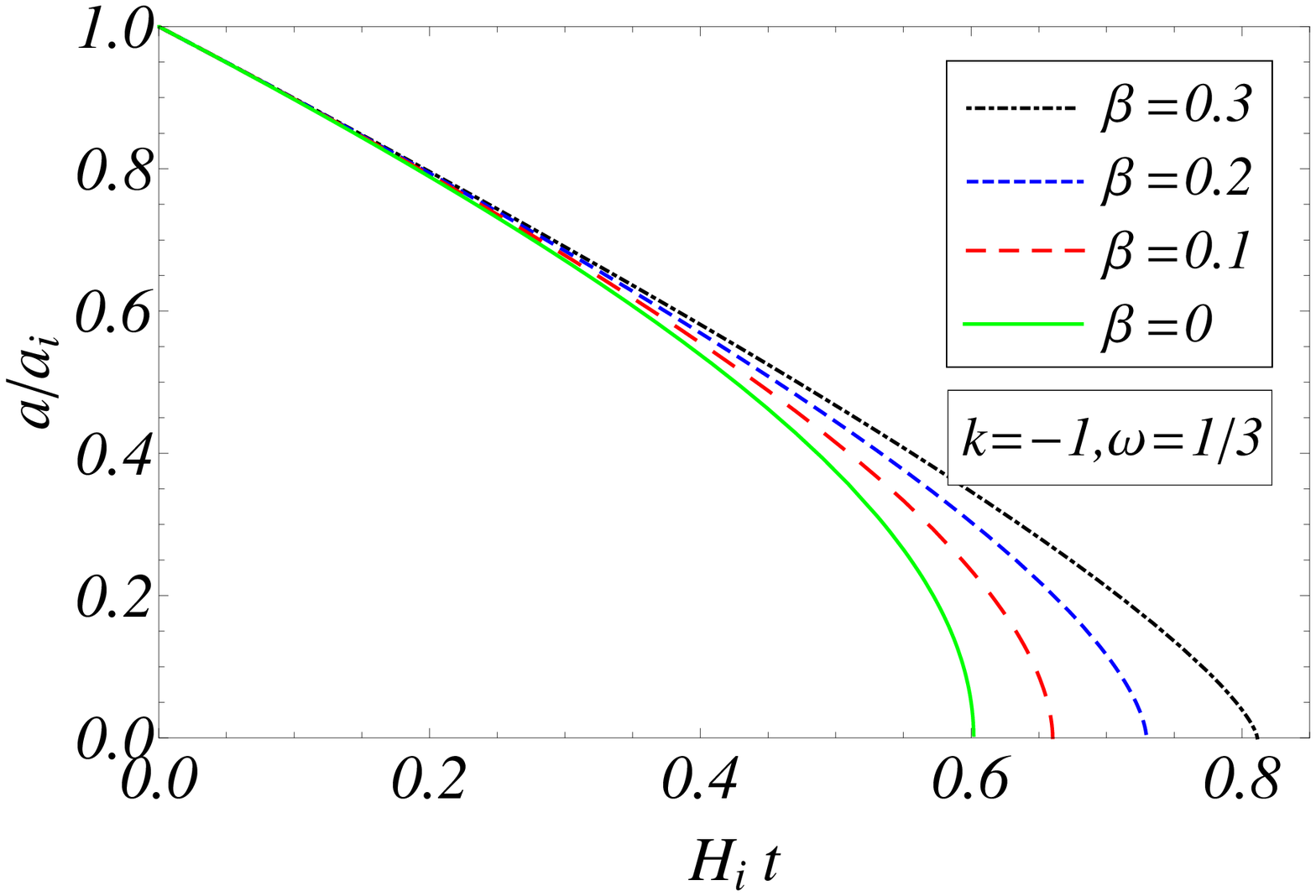,width=2.7truein,height=2.truein}}
                       %\vspace{0.7cm}
                      \caption{Evolution of the scale factor for closed, flat and hyperbolic geometries. The star fluid is represented by a vacuum
component interacting with radiation ($\omega=1/3$). The overall effect of the vacuum component for any geometry is to increase the collapsing time relatively  
to a pure fluid medium ($\beta =0$).  Note also that for a selected value of $\beta$, the collapse process in  closed models ($k=+1$) is favored as compared to the flat case.  In the hyperbolic  geometry ($k=-1$) we have the opposite effect.}
                    \end{figure*} 
\subsection{Basic Equations and Solutions}

In the background given by Eq. (\ref{2.1}),  the EFE  for the interacting mixture (perfect fluid plus a vacuum component) can be written as: 
\begin{eqnarray}
8\pi G \rho_f +\Lambda \left(t \right) &=&3H^2+\frac{3k}{a^2}, \\
8\pi G p_f -\Lambda \left( t \right ) &=&-2\dot{H}-3H^{2}-\frac{k}{a^2},
\end{eqnarray}
where a dot means time derivative, $H=\dot a/a < 0$ is the ``Hubble function'' and $\rho_f$, $p_f$ are, respectively,  
the energy density and pressure of the fluid component which obeys the equation of state 

\begin{equation}\label{EoS}
p_f=\omega \rho_f \,, 
\end{equation}
where  $0\leq\omega\leq 1$ is a constant parameter. The energy conservation law which is is contained in the EFE equations reads: 

\begin{equation}\label{coupling1}
\dot \rho_f + 3\frac{\dot a}{a}\left(\rho_f + p_f\right) \ = - \dot \rho_v.
\end{equation}

For the sake of generality, in what follows  we consider that the  $\Lambda(t)$-term  is given by: 
                \begin{equation}\label{Lambda}
                \Lambda =\Lambda_0 + 3\beta H^{2}+3\beta \frac{k}{a^2},
                \end{equation}
where $\beta$ is a dimensionless constant parameter and the factor 3 was added for mathematical convenience \cite{CLW}. This form is a natural extension of the $\Lambda$-term used in the Paper I (see Eq.(9) there) including the curvature effect. As one may check, it means that the fraction

\begin{equation}
\beta= \frac{\rho_v - \rho_{vo}}{\rho_f + \rho_{v}},
\end{equation}
remains constant during the collapsing process. Such an expression  generalizes the expression derived by Sol\'a and Shapiro \cite{SS}  within a renormalization 
group approach ($k=0$) and also the one proposed by Carvalho {\it et al.} \cite{CLW} by including a bare cosmological constant $\Lambda_0$.

Now, by using expressions (\ref{EoS}) and (\ref{Lambda}), it is easy to check that the differential equation governing the scale factor takes the form: 
              \begin{equation}\label{Eq.0}
              a\ddot{a} + \Delta\, \dot a ^2 +\Delta k-\Lambda_0(1+\omega)a^2/2 = 0,
              \end{equation}
where $\Delta=-1+3(1+\omega)(1-\beta)/2$.
By integrating the above equation we obtain for the first integral: 
     \begin{equation}\label{Eq.1}
     \dot a^2=b\left(\frac{a_i}{a}\right)^{2\Delta} -k +\frac{\Lambda_0a^2}{3(1-\beta)}, 
     \end{equation}
where  $b=(a_iH_i)^2+k-\frac{\Lambda_0a_i ^2}{3(1-\beta)}$ is a constant, and $a_i$, $H_i$ are the initial values 
for the scale factor and the Hubble parameter.  

Since $\Lambda_0$ is very small, once the collapse process has initiated under the pull of the core self gravitation ($a << a_i$) the term containing $\Lambda_0$  can be surely neglected. Hence, in what follows we will retain only the curvature term in order to quantify its effect on the black hole mass  and other physical quantities. 
In this case, the full integration  of the field equations can be  performed by introducing  the auxiliary variable 
                \begin{equation}\label{Eq.2}
                u=\frac{ k}{b}\left(\frac{a}{a_i}\right)^{2\Delta},
                \end{equation}
which transforms Eq.(\ref{Eq.1}) to 
            \begin{equation}\label{Eq.3}
             \dot u =\frac{2b^{1/2}\left({k}/{b}\right)^A}{\left(2A-1\right)a_i\,}u^{1-A}\,\sqrt{1-u}\,,
             \end{equation}
where $A=\frac{1+\Delta}{2\Delta}$.

The inversion of Eq.(\ref{Eq.3}) results
                           \begin{equation}\label{Eq.4}
                           \frac{dt}{du}=\frac{\left(2A-1\right)a_i\,b^{A-\frac{1}{2}}}{2k^A}\frac{u^{A-1}}{\sqrt{1-u}}\,,
                           \end{equation}
and combining with the second derivative we find
                \begin{equation}\label{Eq.5}
                u(1-u)\frac{d^2 t}{d u^2}+\left[ (1-A)-\left(\frac{3}{2}-A\right)u    \right]\frac{dt}{du}=0\,.
                \end{equation}
The above Eq. (\ref{Eq.5}) is a particular case of the hypergeometric  differential equation with parameters $\alpha_1=0$, $\alpha_2=\frac{1}{2}-A$, and $\alpha_3=1-A$, and whose solution is given by \cite{Abra}
            \begin{equation}\label{Eq.6}
             t(u)=c_1+c_2u^A F\left(\frac{1}{2},A;1+A;u\right),
             \end{equation}
where $c_1$ and $c_2$ are arbitrary constants and $F(\alpha_1, \alpha_2, \alpha_3, u)$ is the hypergeometric Gaussian function \cite{Abra}.
Considering the initial condition $a(t=0)=a_i$, and that for $t=t_c \rightarrow a=0$, we can write the above solution as

\begin{equation}
 \label{Eq.7}
\left(1-\frac{t}{t_c}\right)=\left(\frac{a}{a_i}\right)^{1+\Delta}\frac{F\left(\frac{1}{2},A;1+A;\cfrac{k}{b}\, \left(\cfrac{a}{a_i}\right)
^{2\Delta}\right)}{F\left(\frac{1}{2},A;1+A;\cfrac{k }{b}\right)}\,,
\end{equation}
where
\begin{equation}\label{Eq.8}
 t_c=\frac{H_{i}^{-1}F(1/2,A;1+A;\frac{k}{b})}{(1+\Delta)\sqrt{1 + k/a_{i}^{2} H_{i}^{2}}}\,, 
\end{equation}
is the total collapsing time. As one may check, since $F(\alpha_1, \alpha_2, \alpha_3, 0)=1$, where $\alpha_i$ are arbitrary parameters
of the hypergeometric function, we see that for $k\rightarrow 0$ the above expressions assume the simpler forms 
                         \begin{eqnarray}\label{CL}
\frac{a}{a_i}&=& \left(1-\frac{t}{t_c}\right)^{\frac{2}{3(1+\omega)(1-\beta)}} \, , \\
                          t_c&=&\frac{2H_{i}^{-1}}{3(1+\omega)(1-\beta)}\, , \nonumber
                          \end{eqnarray}
which are identical to the solutions  previously obtained by Campos and Lima for the flat case (see Eqs. (12) and (13) of Paper I).

It is worth noticing that for all values of $\beta$ and k,  the modulus of the initial Hubble function ($H_i$) sets the collapsing time scale  to reach the singular point ($a(t_c)=0$), (see Eq. (\ref{Eq.8})). However,  for $\beta= 1$ the collapsing time $t_c \rightarrow \infty$ and, in this case, the spacetime is nonsingular (pure de Sitter vacuum). As in the Paper I, henceforth it will be assumed that the vacuum parameter is restricted on the interval $0 \leq \beta <1$.

\begin{figure*}[th]\label{fdensity}
                 \centerline{\hspace{1cm}\psfig{figure=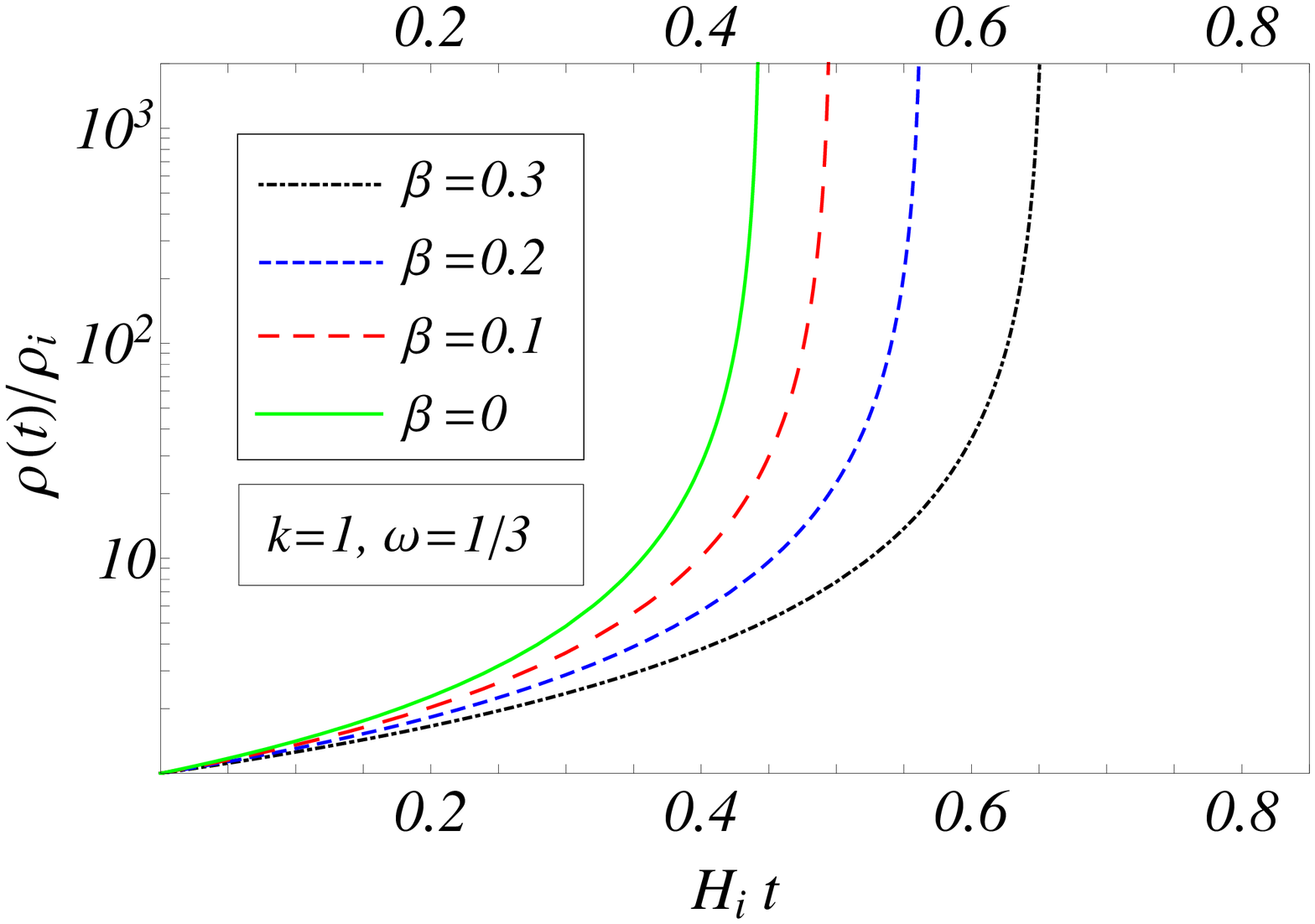,width=2.7truein,height=2.5truein}\hspace{-1.0cm}
								\psfig{figure=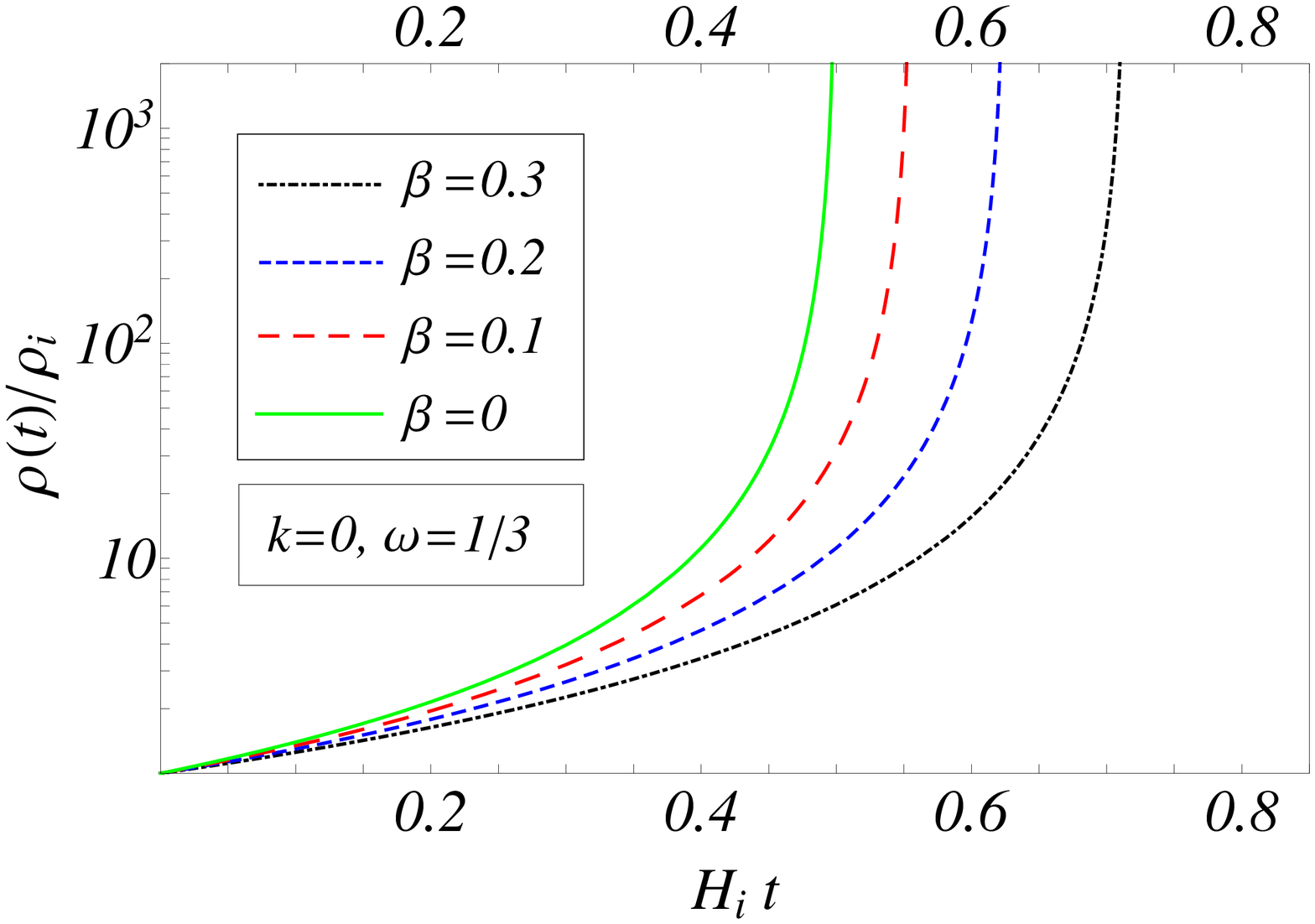,width=2.7truein,height=2.5truein}\hspace{-1.0cm}
                    \psfig{figure=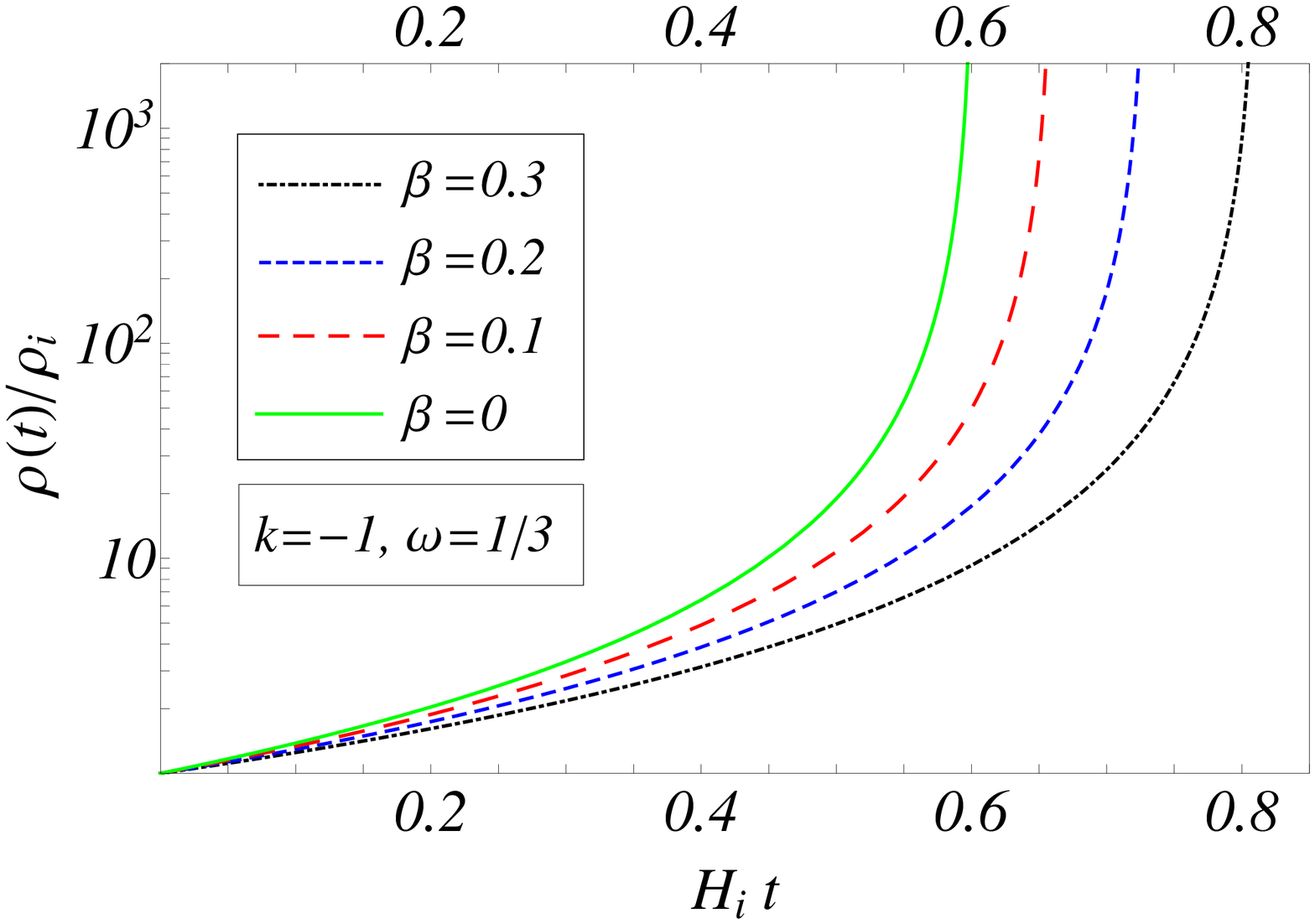,width=2.7truein,height=2.5truein}}
                       %\vspace{0.7cm}
                      \caption{Evolution of the total energy density for  closed, flat and open geometries. For all cases 
the two fluid mixture is formed by a vacuum component interacting with a radiation fluid ($\omega = 1/3$). Note that the evolution of the energy densities
depend on the values of the $\beta$ and curvature parameters.}
                    \end{figure*} 
                    
In Figure 1 we show the evolution of the scale factor as a function of the dimensionless time $H_i t$.
Different values for $\beta$ were selected for three distinct cases: (i) a radiation filled core ($\omega = 1/3$)
coupled to a vacuum growing component and a positive curvature;  (ii) an identical scenario to the anterior case with  null curvature;
(iii) finally, we consider a negative curvature parameter.  Notice that for a fixed curvature parameter, the collapsing time increases for higher values of $\beta$. However, for a given value of $\beta$, models with  positive (negative) curvature  present a smaller (higher) collapsing time relatively to the flat case. For the sake of completeness, we close this section on exact results by exhibiting  the expressions for the vacuum and fluid energy densities:
\begin{eqnarray}
\rho_v&=&\rho_{vi}\left(  \frac{a_i}{a} \right)^{3(1+\omega)(1-\beta)}\, ,\\
\rho_f&=&\rho_{fi}\left(  \frac{a_i}{a} \right)^{3(1+\omega)(1-\beta)}\, ,
\end{eqnarray}
where the initial densities, $\rho_{vi}$, $\rho_{fi}$, are related by the expression $\rho_{vi}=\frac{\beta}{1-\beta}\rho_{fi}$.

In Figure 2, we show the time behavior of the total energy density as a function of the dimensionless time parameter $H_i t$ and some selected values of the pair of free parameters ($\beta$, k).  For all values of  $\beta < 1$, we see that the energy density diverges at the collapse time ($t_c$) which is strongly correlated with  
the values of $\beta$.  As should also be expected (see Fig. 1) for a given value of $\beta$, the  collapsing time is reduced (increased) for positive (negative)  curvatures
as compared to the flat scenario.  In a more realistic treatment, a continuous transition from radiation ($\omega=1/3$) to
to the Zeldovich's stiff-matter fluid  ($\omega =1$) may also occur at the late stages. 
Naturally, such a final state is included in the general solutions for $a(t)$ and $\rho(t)$ with similar plots appearing in Figs. 1 and 2. 
\subsection{Unified conformal time solutions}
Some gain in simplicity is obtained by using the conformal transformation, $dt=a(\eta)d\eta$. In this case,  
the metric given by Eq.(\ref{2.1}) becomes
\begin{equation}
 ds^2= a(\eta)^2\left( d\eta^2-\frac{dr^{2}}{{1-k\,r^2}} - r^{2} d\Omega^{2}  \right)\, ,
\end{equation}
while the motion equation \eqref{Eq.0} is transformed to:
            \begin{equation}\label{Eq.18}
             a a''+(\Delta-1)a'^2+k\Delta a^2=0\,.
             \end{equation}
Where the prime denotes conformal time derivative.
Now, by introducing the auxiliary transformation $z=a^{\Delta}$,
Eq.(\ref{Eq.0}) takes the following forms (see \cite{AL88,Lima01} in the case $\beta =0$):
                \begin{equation}\label{Eq.20}
                 z^{\prime \prime}+k\Delta ^2 z=0\, .
                  \end{equation}
This is a remarkable result for the interacting mixture. We see that the auxiliary scale factor  obeys the same differential equation of a particle subject to a linear force regardless the values of $\beta$ and $\omega$.  In particular, the flat case behaves like a free particle while for a closed geometry ($k=1$) we have an harmonic oscillator with angular ``frequency"  $f =  |-1 + 3(1+\omega)(1-\beta)/2|$.     

A simple integration of above equation yields:
              \begin{equation}\label{Eq.21}
               a(\eta)=\left[c_1\sin\left(\eta \Delta \sqrt{k }\right)+c_2\cos\left(\eta \Delta \sqrt{k }\right)\right]^{1/\Delta}\,,
                \end{equation}
where $c_1$ and $c_2$ are arbitrary constants with the proviso that the first integral given by Eq.(\ref{Eq.1})
must be obeyed, and we have used the inverse transformation $a=z^{\frac{1}{\Delta}}$.

With suitable initial conditions for the collapse process, the above solution can be rewritten as:
\begin{equation}
\label{Eq.22}
a(\eta)=a_i\left\lbrace  \sqrt{\frac{b}{k}}\sin{\sqrt{k}\Delta(\eta_c-\eta)}  
 \right\rbrace ^{\frac{1}{\Delta}}\, ,
\end{equation}
where
\begin{equation}
\label{Eq.23}
 \eta_c=\frac{1}{\Delta \sqrt{k}}\sin ^{-1}\left[ \sqrt{\frac{k}{b}}  \right]\, ,
\end{equation}
is the collapsing time in the conformal coordinate.  
Inserting expression (\ref{Eq.22}) into  Eq.(\ref{Eq.7})
we obtain the general solution for the collapsing time, $t(\eta)$, regardless of the equation of state
and curvature parameters. 

The hypergeometric function can be reduced to an elementary expression for  specific values of the free parameters. In this case,  the conformal solutions are readily obtained from  the general expressions of a($\eta$) and t($\eta$). In order to exemplify that, let us now exhibit the explicit solutions describing the last stages of the collapsing core by assuming the Zeldovich stiff matter fluid  ($\omega =1$) plus a vacuum component ($\beta = 1/2$) for all values of the curvature parameter.\\
 
\begin{figure*}[th]\label{ftime}
\centerline{\hspace{1cm}\psfig{figure=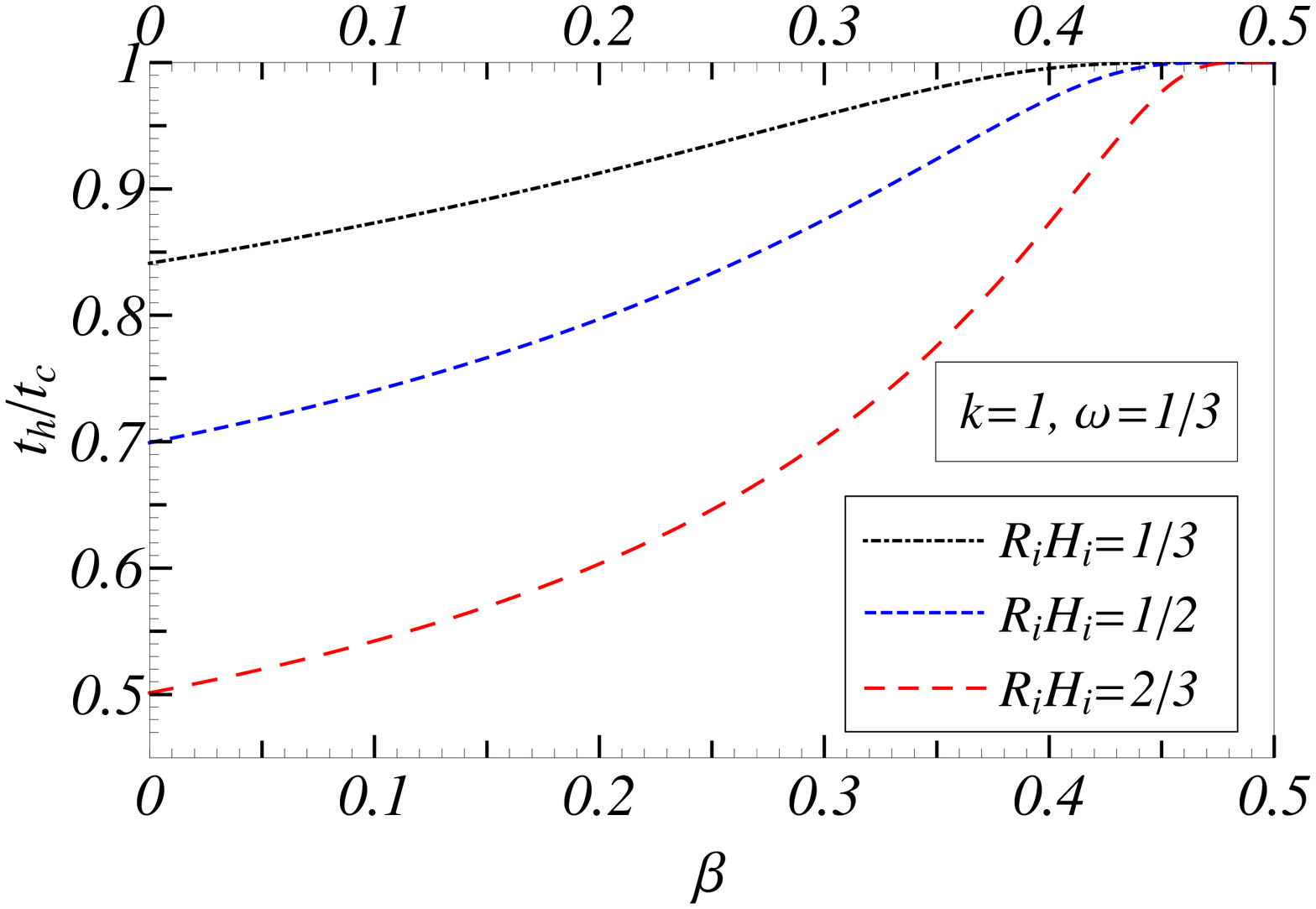,width=2.7truein,height=2.2truein}\hspace{-1.0cm}
                                          \psfig{figure=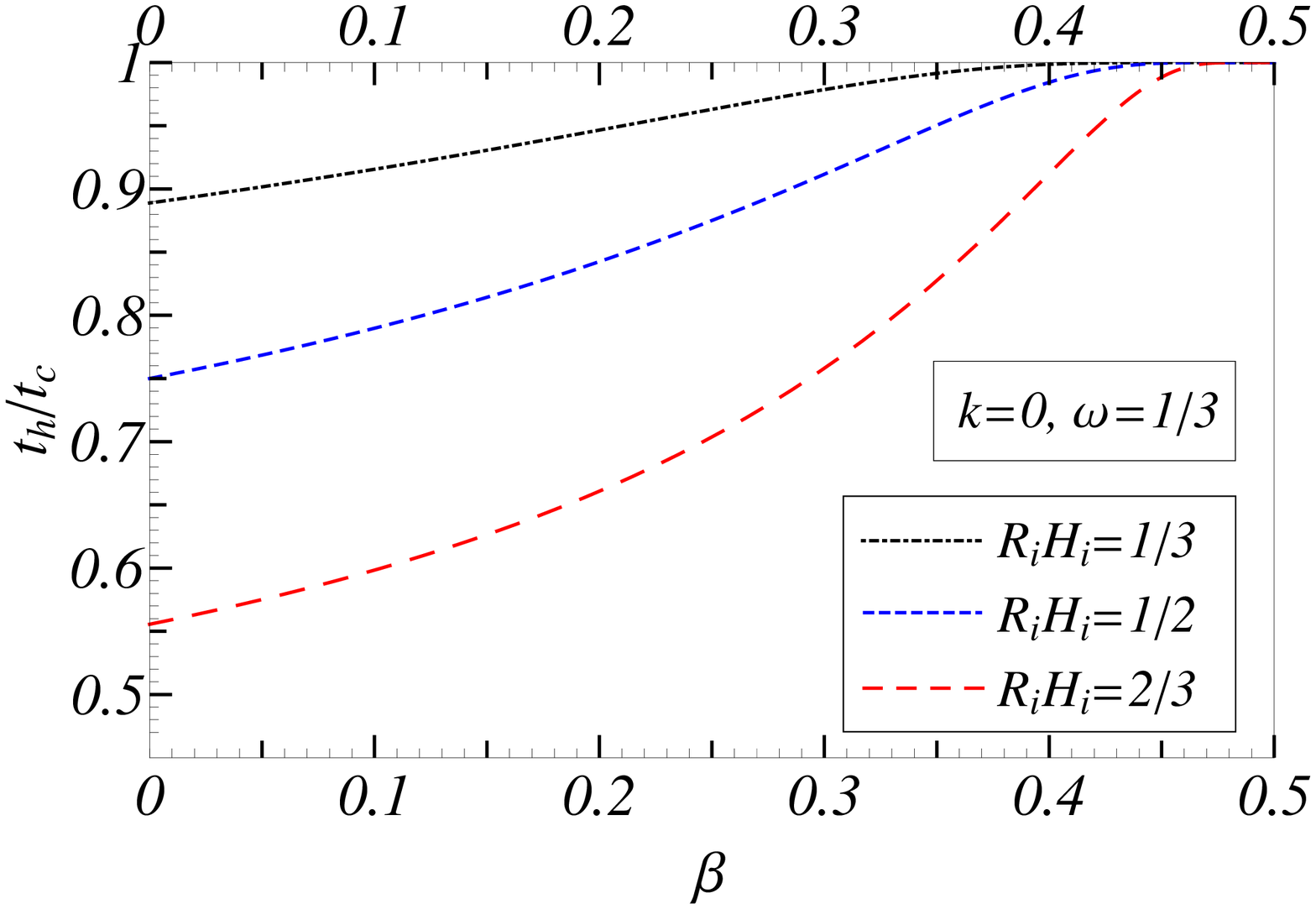,width=2.7truein,height=2.2truein}\hspace{-1.0cm}
                                             \psfig{figure=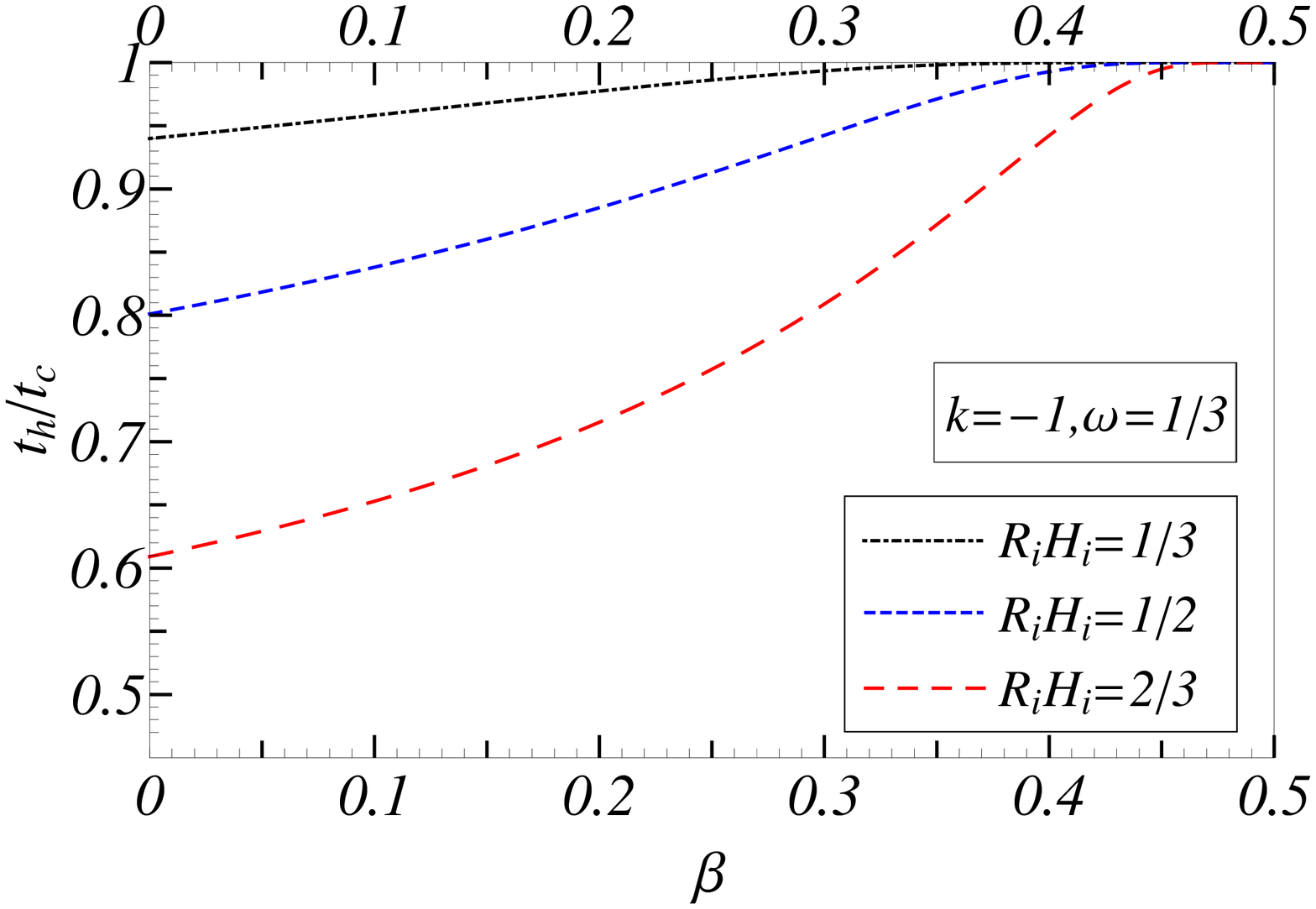,width=3truein,height=2.2truein}}
                       %\vspace{0.7cm}
                       
\caption{Ratio between the the time to develop the apparent horizon ($t_{AH}$)  and collapsing time ($t_{c}$).  The plots show the case  of a radiation fluid for  closed, flat and hyperbolic geometries.  We see that the formation of the horizon is delayed for higher values of $\beta$. Note that the for a given value of $\beta$ the formation of the apparent horizon  (as compared to the collapsing time) is slightly favored for closed models ($k=1$), and the inverse happens in the hyperbolic case ($k=-1$).  We also remark that for $\beta > 0.5$ the apparent horizon is not formed before the total collapsing time, and, therefore, the singularity is naked.}                          
\end{figure*} 
 
Let us first determine the unified form of the solutions. For $\omega =1$ and $\beta = 1/2$ it follows that $\Delta = 1/2$. In this case, from  Eqs. (\ref{Eq.7}) and (\ref{Eq.22}), it is easy to see that the unified solution (for any curvature)  can be written as:

\begin{eqnarray}
a(\eta)&=&a_i\,b\left\lbrace  \sqrt{\frac{1}{k}}\sin\left[\sqrt{k}\,\frac{\eta_c-\eta}{2}\right]  
 \right\rbrace ^{2},\\
t_c-t&=&\frac{a_i\,b}{2\,k}\left\{\eta_c-\eta-\frac{\sin\left[\sqrt{k}(\eta_c-\eta)\right]}{\sqrt{k}}\right\},\,
\end{eqnarray}
where $\eta_c$ is given by \eqref{Eq.23} with $\Delta=1/2$. Note also that the parameter $b=a_i^{2}H_i^{2} + k$ is still free to be fixed. By choosing $b=1$ we see that $H_i^{2} = (1-k)/a_i^{2}$. Therefore, for $k=1$ we have $H_i=0$ and the collapse process may start from the rest. In the simplified forms below we have fixed  $b=1$ for all cases.

(i) Closed Solution:  
\begin{eqnarray}
 a(\eta)&=&\frac{a_i}{2}(1+\cos\eta)\, , \nonumber \\
t_c - t&=&\frac{a_i}{2}(\pi-\eta -\sin\eta) \, , \nonumber
\end{eqnarray}
where the collapsing time is $t_c = a_i\eta_c/2$ and $\eta_c = \pi$, according to \eqref{Eq.8} and \eqref{Eq.23}.\\

(ii) Hyperbolic Solution:
\begin{eqnarray}
 a(\eta) &=& \frac{a_i}{2}\left[\cosh(\eta_c-\eta) -1 \right]\, , \\
t_c -t &=& \frac{a_i}{2}\left[ \sinh(\eta_c-\eta) -(\eta_c-\eta) \right]\, ,
\end{eqnarray}
where $\eta_c=2\ln(1+\sqrt{2})$, and  $t_c=a_i(\sqrt{2}-\eta_c/2)$ is given by \eqref{Eq.8} and \eqref{Eq.23}.\\

(iii) Flat solution: 
\begin{eqnarray}
 1-\frac{t}{t_c}&=&\left(\frac{\eta_c - \eta}{2}\right)^3\, , \\
a(\eta)&=&a_i\left(\frac{\eta_c - \eta}{2}\right)^2 \,,
\end{eqnarray}
where $\eta_c=2$, $t_c=2a_i/3$, and $a_i = H_i^{-1}$. Combining the above  expressions one obtains the simplified form: 
\begin{eqnarray}
1-\frac{t}{t_c}=\left(\frac{a}{a_i}\right)^{3/2} \, , \nonumber
\end{eqnarray}
as obtained before through a different method (see Eqs. (\ref{CL})).  

\section{Black  hole formation and apparent horizons}

An important issue on black hole physics is related to the formation 
of an apparent horizon during the collapsing process. In the present context such a subject deserves a closer scrutiny since the collapsing mixture involves a growing vacuum component. 

On the other hand, by the cosmic censorship hypothesis, singular points must be dressed by apparent horizons i.e. naked singularities are forbidden \cite{RP69}.  This means that when the system is approaching the singularity, an apparent horizon  should be developed inside it (before the collapsing time $t_c$). 

Nevertheless, such a conjecture  is not true in general \cite{Papa} and some examples have already been discussed in the literature \cite{Collapse}. In the  present case, the increasing negative pressure of the vacuum component can be responsible by the formation of a naked singularity. 
As we shall see next, for a given interval of the $\beta$ parameter, naked singularities can generically be  formed for arbitrary values of the curvature. 

%Let us now consider a two-sphere $S$ embedded in a slice $\Sigma$ of the spacetime endowed with an outward unit spacelike vector   
%$s^\mu$ and a future-pointing timelike
%unity normal to $\Sigma$ $n^\mu$ the.  Hence, the vector $k ^\mu = s^\mu +n^\mu$ is a nulll vector, and $S$ is a marginally trapped 
%surface if $k ^\mu _{;\mu} = 0$ holds everywhere on the $S$ \cite{Anninos}.%%

%Although considering that the matching conditions 
%and the spacetime outside the star will affect the total mass and the global structure of the 
%black hole,   In other words, the ultimate formation of the black hole singularity does not depend neither on %
%the matching nor the choice of the spacetime outside the star. Here we are mainly interested to discuss under which 
%conditions  a BH is formed during the collapse of  a fluid with a growing interacting vacuum component 
%with arbitrary curvature parameter.

The natural observers describing the behavior of the matter fields in the FLRW geometry are comoving with the fluid volume elements. Let us now 
define a constant geometrical radius ($r_\Sigma $) for the surface dividing the star interior from the exterior, 
For such surface,  the metric can be written as:
\begin{equation}\label{Eq.9}
ds^2 _\Sigma = d\tau ^2 -R\left(\tau\right) ^2 d\Omega ^2 \, ,
\end{equation}
where $t=\tau $ and $R(t) = r_\Sigma a(\tau)$. Apparent horizons are defined by space-like surfaces
with future point converging null geodesics on both sides of the surface \cite{Hawking,Anninos}.  Naturally, for an initially 
untrapped star, the black hole is formed only when the apparent horizon appears before the singularity otherwise a
naked singularity it will be the final stage of the collapsing core.

As discussed by many authors \cite{McV,CM73,CaiWang}, the formation of the apparent horizon is determined  by the condition:
\begin{equation}\label{Eq.10}
R_{,\alpha} R_{,\beta}g^{\alpha \beta}= \left( r_\Sigma \dot a \right) ^2 -1+kr_\Sigma^2 =0 \, ,
\end{equation}
where $()_{,x}=\frac{\partial }{\partial x}$ and $R(t,r)=a(t)r$.

Since the star is initially not trapped the comoving surface is spacelike. This means that   
\begin{equation}\label{Eq.11}
R_{,\alpha}R_{,\beta}g^{\alpha \beta}=\left[ r \dot{a}\left( t_i \right) \right] ^2 -1+kr^2 < 0\, ,
\end{equation}
which implies that $0<(R_iH_i)^2<1-kr^2$. 

Another important quantity is the mass function that furnish the total mass inside  the surface with radius $r$ at time $t$. 
Cahill and McVittie \cite{CM73} wrote such a function for a particular reference system that here takes
the following form (see also \cite{Poisson})
\begin{equation}\label{Eq.12}
m(t,r)=\frac{1}{2}R\left( 1+R_{,\alpha}R_{,\beta}g^{\alpha \beta}\right) =\frac{1}{2}R (\dot{R}^2+kr^2) .
\end{equation}

As it appears, the first integral given by equation \eqref{Eq.1} can be rewritten in the variable $R =a(t)r$ by taking into account the 
constraint to form the apparent horizon
\begin{equation}\label{Eq.13}
 \dot R_{AH}^2 + kr^{2}= b\,r ^2\left(\frac{R_i}{R_{AH}}\right)^{2\Delta}=1\, ,
\end{equation}
and solving for the radius of the apparent horizon we obtain
\begin{equation}\label{Eq.14}
 \frac{R_{AH}}{R_i}=\left(R_i^2H_i^2+kr^2\right)^{\frac{1}{2\Delta}}\, .
\end{equation}
On the other hand, at $t=t_{AH}$, the time evolution equation (\ref{Eq.7}) implies that
\begin{equation}\label{Eq.14a}
\frac{t_{AH}}{t_c}= 1 - \left(\frac{R_{AH}}{R_i}\right)^{1+\Delta}
\frac{F(1/2,A;1+A;\frac{k}{b}(\frac{R_{AH}}{R_i})^{2\Delta})}
{F(1/2,A;1+A;\frac{k}{b})}.
\end{equation}

Notice that for $k=0$ the above expression  reduce to
%\begin{equation}\label{Eq.15}
%a_{AH}=a_i(R_i\,H_i)^{1/\Delta_1} \nonumber \, ,
%\end{equation} 
%and
\begin{equation} \label{Eq.16}
t_{AH}/t_c = 1 - (R_i\,H_i)^{\frac{1+\Delta}{\Delta}}\,,
\end{equation}
which is exactly the same one previously obtained in Paper I with a slightly different notation (see Eq. (26) there).

In Figure 3, we show the behavior of the dimensionless ratio $t_{AH}/t_c$  as a function of $\beta$ parameter for a collapsing radiation fluid ($\omega = 1/3$) in different geometries ($k = 0, \pm 1$).  For a selected  value of $R_i H_i$, we see that the curvature effects are small in comparison to the flat case. Note also  that in the examples displayed the formation of the apparent horizon happens before the emergence of the singularity. 

\begin{figure*}[th]\label{fblackmass}
                 \centerline{\hspace{0.5cm}\psfig{figure=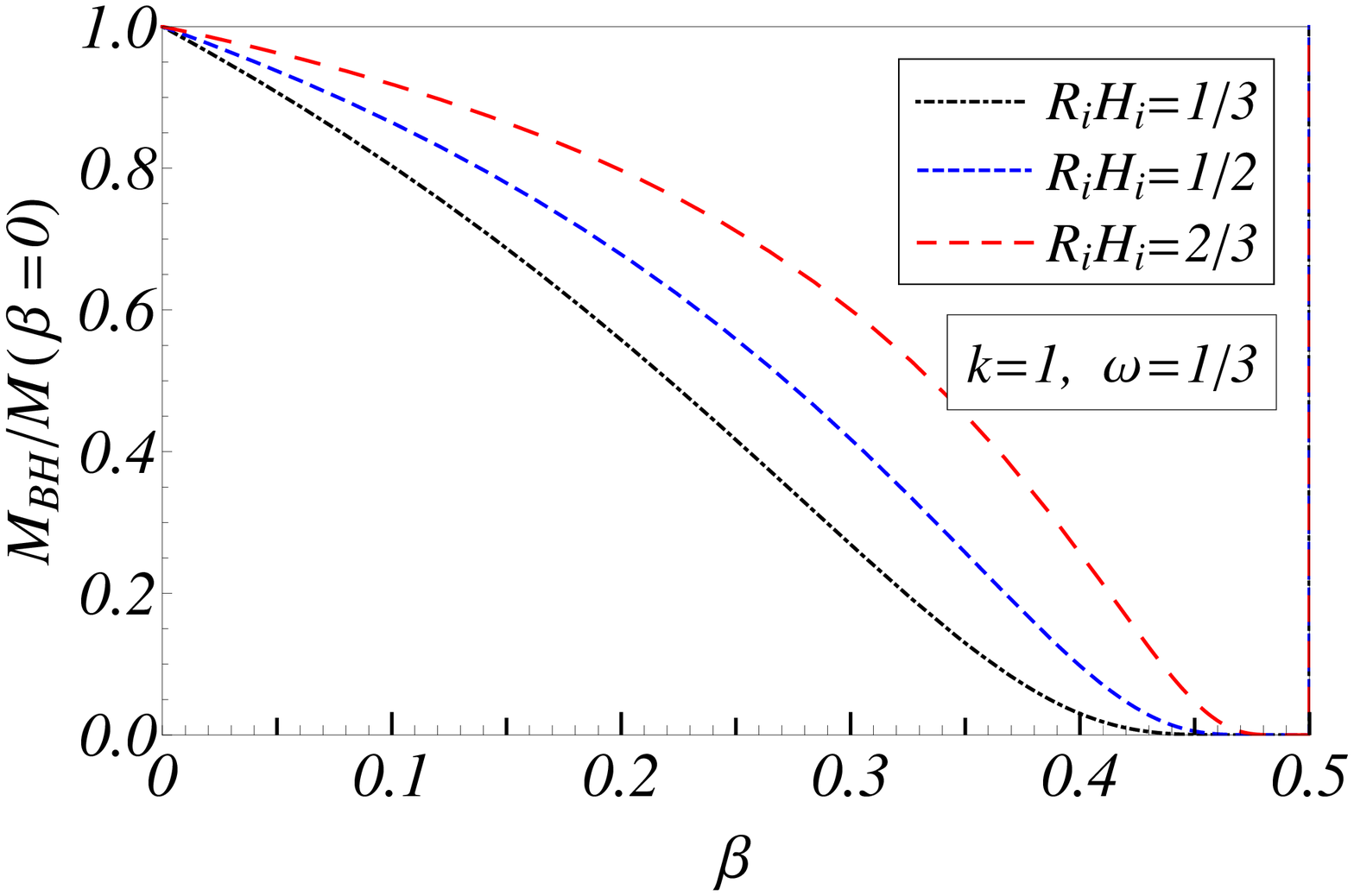,width=2.7truein,height=2.truein}\hspace{-1cm}
								\psfig{figure=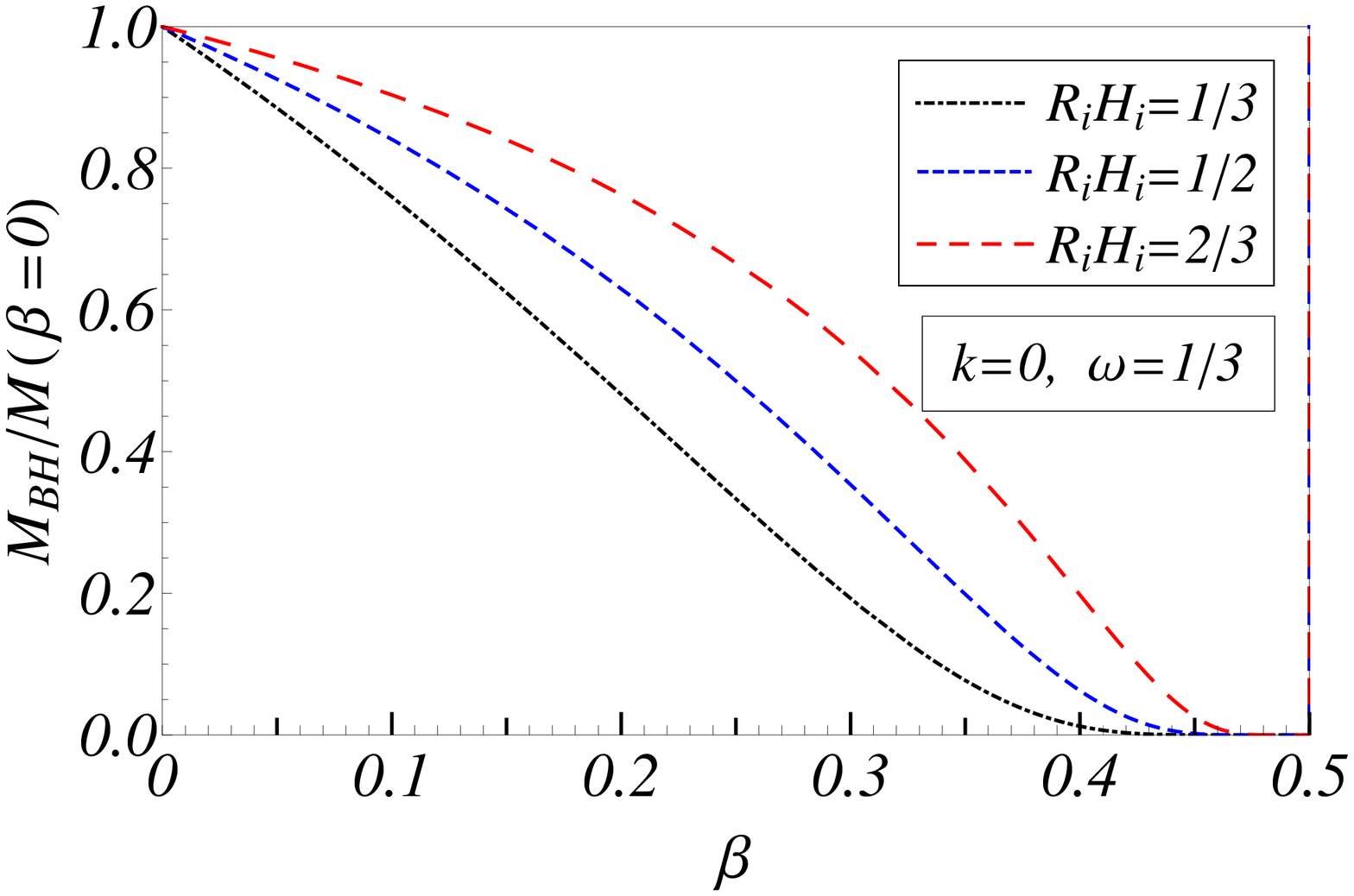,width=2.7truein,height=2.truein}\hspace{-1cm}
                    \psfig{figure=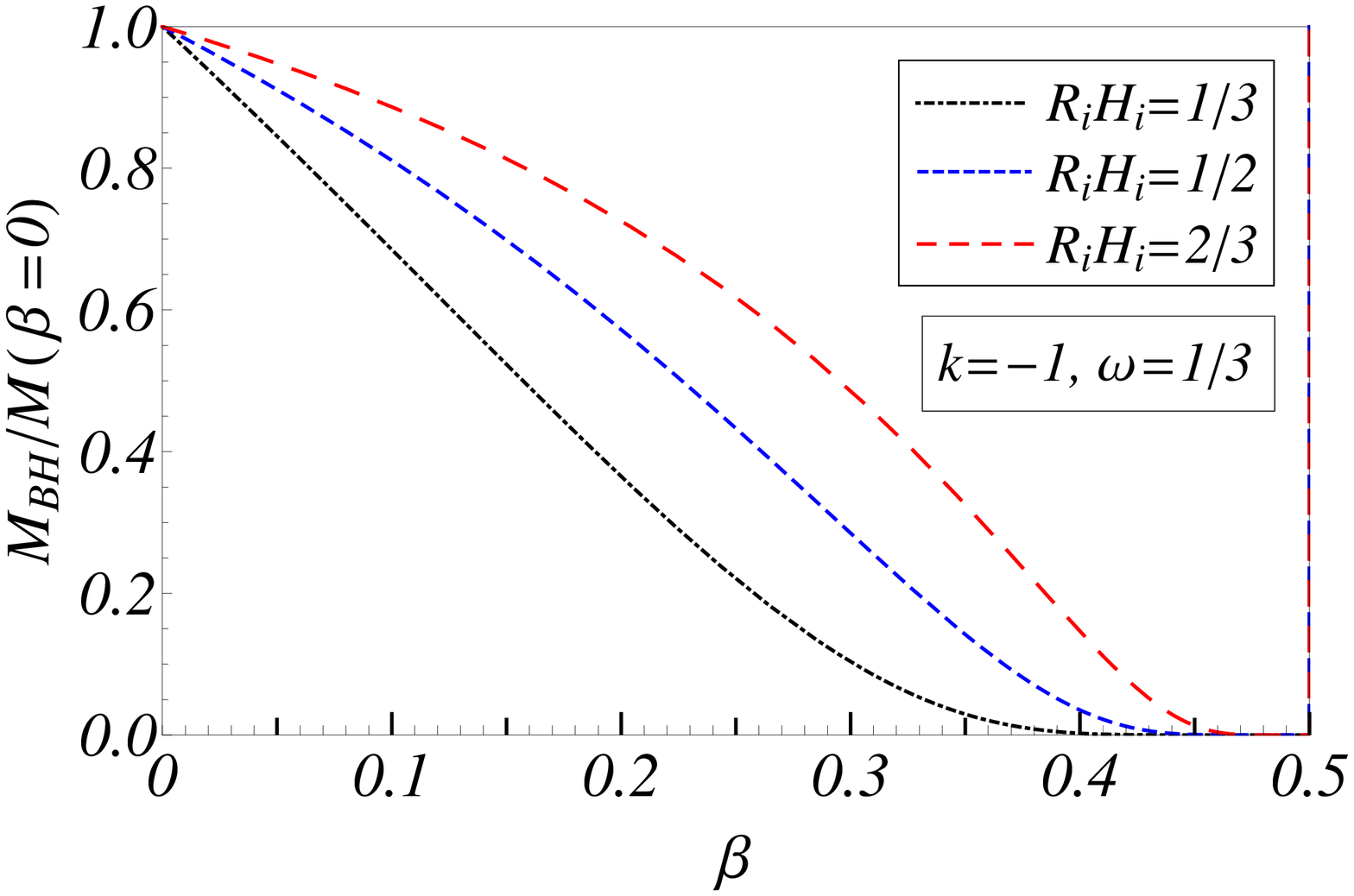,width=2.7truein,height=2.truein}}
                      \caption{Black hole masses for closed, flat and hyperbolic geometries. 
{ The ratio between the mass of the BH formed and the mass of a pure fluid versus the $\beta$ parameter.}  
  For a fixed value of the product $H_iR_i$, we see that  
the BH mass is not only heavily dependent on the $\beta$ parameter but decreases for higher values of
 $\beta$. For $k=+1,0,-1$ the mass of the black hole formed diminishes, respectively.  For all geometries we have adopted  the Cahill and McVittie mass definition \cite{CM73}.}
                    \end{figure*} 

Following the same steps of paper I and using the mass definition given by Eq. (\ref{Eq.12}) we can calculate the total mass of the black hole: 
\begin{equation}
 M(\tau_{AH})=M_0\left\lbrace  \frac{R_i^2H_i^2+kr^2}{(R_iH_i)^{4\Delta}}  \right\rbrace ^{\frac{1}{2\Delta}}\, ,
\end{equation}
where $M_0={R_i^3H_i^2}/{2}$ is the black hole mass for $k=\omega=0$.
For a flat geometry ($k=0$)  the above expression reduces to
\begin{equation}
M(\tau _{AH})=\frac{1}{2}R_i^3H_i^2(R_i H_i)^{\frac{3(1+\omega)(1-\beta)-3}{1-\frac{3}{2}(1+\omega)(1-\beta)}} \, ,
\end{equation}
which is identical to the expression  previously obtained in Paper I (cf. Eq.(29) there).

%{\bf To have a better idea of the collapsed mass we can expressed it in solar mass units.  However, it shoul be stressed that in our study
%the final fate of the collapse process are formed from collapsing star cores.  Consequently, supermassive black holes and the 
%recently discovered quasar at $z=7.085$ with $2\times 10^9 M_\odot $ \cite{Mortlock} are not related with the framework in this study.  Generally, the large mass are the result %of the cosmological accretion mechanisms.

In terms of solar mass units such an expression can be rewritten as:
\begin{equation}\label{mass}
 M(\tau_{AH})=R_i\frac{M_\odot}{r_{S\odot}}\left\lbrace R_iH_i\sqrt{\Omega_{fi}/(1-\beta)}\right \rbrace ^{\frac{1}{\Delta}} \, ,
\end{equation}
where $M_{\odot}$ and $r_{S\odot}$ are respectively, the solar mass and the Schwarzschild radius of the sun.
In the case of a fluid with $k=0$  one finds $\Omega_{fi} = 1 - \beta$  with the above expression reducing to 
\begin{equation}
M(\tau _{AH})= \frac{R_i M_{\odot}}{r_{S\odot}} (R_i H_i)^{\frac{2}{{{3}(1+\omega)(1-\beta)} - 2}}.
\end{equation}
Now, assuming $R_i = 80\, Km$ and $R_iH_i \sim 1/2$  one finds 
$M \sim 13.3 \, M_{\odot}$ for a pure radiation fluid ($\beta=0, \omega = 1/3$). However, for  $\beta = 0.25$ and the same values of $R_i$ and $H_i$, one obtains a smaller mass, namely, $M \sim 6.6 \, M_{\odot}$.   Such results reproduce the values obtained in the paper I where $R_i$ was wrongly typed.  

What about the influence of the curvature on the black hole mass? Taking into account identical conditions with a non null curvature is not difficult to conclude  that the influence of the curvature is to mildly increase (decreases) the black hole mass for closed ($k=1$) and hyperbolic geometries, respectively.   Since the black hole mass is a positive quantity,  we  can also infer from (\ref{mass}) that the vacuum parameter must satisfy 
\begin{equation}\label{Eq.17}
\beta \leq 1-\frac{2}{3(1+w)}, 
\end{equation}
in order to form a black hole. For $\omega=1/3$ this inequality implies that $\beta \leq 0.5$  in agreement with the results presented in Figs. 3 and 4.  Naturally, for $\beta > 0.5$ the singularity of the fluid mixture (radiation plus vacuum) is naked. 
\section{Final Comments}
In this paper we have discussed the collapsing process of a massive star core whose matter content is formed by a mixture of two interacting components:  a simple fluid described by the EoS, $p=\omega \rho$, $\omega$=constant, and a growing vacuum component. Through an unified  treatment  we have taken into account the curvature effects.  The vacuum energy density was supposed to satisfy the constraint $\rho_v = \beta \rho_{total}$ where $\beta$ is a constant. From the FLRW equations this means that the  dynamical $\Lambda(t)$-term
satisfies the relation \cite{CLW,SS}: $\Lambda(H,a)$ = $3\beta H^{2}+3\beta k/{a^2}$, where $k=0, \pm 1$. The fluid component satisfies all the energy conditions (weak, dominant and strong), but the vacuum component ($p_v= -\rho_v$)  violates the strong energy condition (SEC).  

The main results derived here can be summarized in the following statements:

(i) For all values of the curvature,  the vacuum component though violating the SEC does not avoid the ultimate collapse of the massive core, i.e. the system reach the singular point  for a broad range of the free parameters ($\omega$, $\beta$). 

(ii) For a fixed value of the initial conditions and free parameters, we shown that the collapsing process  is slightly favored  for  closed  geometry  relatively  to the flat case. For negative curvature we have found the opposite effect. 

(iii) The time to form the apparent horizon ($t_{AH}$) and the total collapsing time ($t_c$) were analytically obtained through a unique expression valid for all physical values of the free parameters $\omega$, $\beta$ and k (see Eq.(2.41)).  More interesting, by using the ratio $ t_{AH}/t_c$  was also possible to determine when black holes or  naked singularities are formed. 

(iv) The nature of the singularity is heavily dependent on the values of the $\beta$ parameter. For  $\beta < 1 - 2/3(1 + \omega)$  a black hole is formed, however, if $\beta > 1 - 2/3(1 + \omega)$ an apparent horizon does not appear before the collapsing time $t_c$, and, therefore, the cosmic censorship conjecture is violated  and a naked singularity is formed. The critical case,  $\beta = 1 - 2/3(1 + \omega)$, defines the boundary between naked singularities and black holes. We stress that such a condition  does not depend on the curvature.
Th existence of naked singularities  is  related to the vacuum pressure, and, generically, it appears  when 
negative pressures large enough take place in the mixture.  

%Naturally, all the results derived here are heavily dependent on the  form assumed for $\Lambda(t)$ and even the avoidance of 
%the singularity may occur whether a more realistic description of the decaying vacuum is adopted. Nonetheless, the results obtained
%here suggest that a growing vacuum energy density may lead at the late stages to naked singularities even when inhomogeneities are taken into account. 
%The physical  effects of a growing time varying vacuum  energy density on the inhomogeneous collapse will be discussed in a forthcoming communication.  

Finally, we emphasize that the main caveat of the model discussed here is related to the  hypothesis of homogeneity and isotropy of the spacetime generated by the collapsing massive star core. Presumably, it must be approximately valid at the late stages of the collapse.  Note also that the avoidance of the singularity was not possible even in the presence of the vacuum energy density, but, clearly,  such a prediction  can be only an artifact of the decaying model assumed in the present paper. The generality of such  a result will be discussed in a forthcoming communication.  

%is the meansince the pressure should be made to vanish at the boundary
%thereby obtaining a smooth matching with the Schwarzschild-de Sitter vacuum solution outside the star. 
%
%
%
\section{Acknowledgments}
This paper is dedicated to  the late Professor Mario Jos\'e Delgado
Assad  (UFPB-Brazil).
His earlier contribution  to the unified approach for FRW type
cosmologies  motivated the present work.
The authors E.L.D.P. and M.C. are partially supported by a grant from
CNPq and J.A.S.L. is partially supported by  CNPq and FAPESP (No.
04/13668-0).


\begin{thebibliography}{100}

\bibitem{Riess} A. Riess {\it et al.}, Astron. J. {\bf{116}}, 1009 (1998);
S. Perlmutter {\it et al.}, Nature, {\bf{ 391}}, 51 (1998);
M. Kowalski {\it et al.}, Astrophys. J. {\bf{686}}, 749 (2008);
R. Amanullah {\it et al.}, Astrophys. J. {\bf{716}}, 712 (2010). 

\bibitem{Komatsu} D.~N.~Spergel {\it et al.}, {Astrophys.\ J.\ Suppl.} {\bf 148}, 175 (2003);
D. N. Spergel {\it et al.} Astrophys. J. Suppl. Ser. {\bf 170}, 377 (2007);
E. Komatsu {\it et al.}  Astrophys. J. {\bf 192}, 18 (2011). 

\bibitem{rev1} P. J. E. Peebles and B. Ratra, Rev.~Mod.~Phys. {\bf 75} 559 (2003);
T. Padmanabhan, Phys.~Rept. {\bf 380}, 235 (2003);
J.~A.~S. Lima, Braz.~Journ.~Phys. {\bf 34}, 194 (2004), astro-ph/0402109;
E. J. M. Copeland and ~S. Tsujikawa, Int.~J.~Mod.~Phys. {\bf D15}, 1753 (2006);
J. A. Frieman, M. S. Turner and D. Huterer,  Ann.~Rev.~Astron. \& Astrophys, {\bf 46}, 385 (2008);
M. Li {\it et al.},  arXiv:1103.5870 (2011).

\bibitem{Zeldovich67} Ya. B. Zeldovich, Usp. Fiz. Nauk {\bf 94}, 209 (1968) [Sov. Phys. Usp. 11, 381 (1968).

\bibitem{zee85} A. Zee, in High Energy Physics, Proceedings of the 20th
Annual Orbis Scientiae, edited by B. Kursunoglu, S. L.
Mintz, and A. Perlmutter (Plenum, New York, 1985).

\bibitem{Weinberg}  S. Weinberg, {Rev. Mod. Phys.} {\bf 61}, 1 (1989).


%\bibitem {Bas2010} S. Basilakos and J. A. S. Lima, Phys. Rev. D {\bf 82}, 023504 (2010), [arXiv:1003.5754]; 
%S. Basilakos, M. Plionis and J. A. S. Lima Phys. Rev. D {\bf 82}, 083517 (2010), arXiv:1103.1464 [astro-ph.CO]

\bibitem{sources} B. Ratra and P. J. E. Peebles, Phys. Rev. D {\bf 37}, 3406 (1988);
C. Wetterich, Astron. and Astrophys. {\bf 301}, 321 (1995);
P. G. Ferreira and M. Joyce, Phys. Rev. D {\bf 58}, 023503 (1998);
N. J. Poplawski, [arXiv:gr-qc/0608031v2] (2006). 

\bibitem{OT86} M. Ozer and M. O. Taha, Phys. Lett.  B {\bf 171}, 363 (1986);
Nucl. Phys. B {\bf 287} 776 (1987).

\bibitem{L1} K. Freese {\it et al.}, {Nucl. Phys. } B {\bf 287}, 797 (1987);  
W. Chen and Y-S. Wu, {Phys. Rev.} D {\bf 41}, 695 (1990);
D. Pav\'on, {Phys. Rev.} D {\bf 43}, 375 (1991).

\bibitem{CLW} J. C. Carvalho, J. A. S. Lima, and I. Waga, Phys. Rev.  D {\bf 46}, 2404 (1992).

\bibitem {L2} J. A. S. Lima and J. M. F. Maia, {Phys. Rev.} D {\bf 49},  5597 (1994);  
J. A. S. Lima and M. Trodden, Phys. Rev. D {\bf 53}, 4280 (1996),  astro-ph/9508049.

\bibitem{L3} I. Waga, Astrophys. J. {\bf 414}, 436 (1993);
L. F. Bloomfield Torres and I. Waga, Mon. Not. R. Astron. Soc. {\bf 279}, 712 (1996);
A. I. Arbab and A. M. M. Abdel-Rahman, {Phys. Rev. } D {\bf 50}, 7725 (1994);
J. M. Overduin and F. I. Cooperstock, Phys. Rev.  D {\bf 58}, 043506 (1998); 


\bibitem{L5} R. G. Vishwakarma, Class. Quant. Grav. {\bf 17}, 3833 (2000);
M. V. John and K. B. Joseph, Phys. Rev. D {\bf 61}, 087304 (2000);
O. Bertolami and P. J. Martins, Phys. Rev. D {\bf 61}, 064007 (2000);
R. G. Vishwakarma, Class. Quant. Grav. {\bf 18}, 1159 (2001);
A. S. Al-Rawaf, {Mod. Phys. Lett.} A {\bf 16}, 633 (2001). 


\bibitem{new1} M. K. Mak, J. A. Belinchon, and T. Harko, Int. J. Mod. Phys. D {\bf 14}, 1265 (2002);
M. R. Mbonye, Int. J. Mod. Phys.  A {\bf 18}, 811 (2003);
J. V. Cunha and R. C. Santos, Int. J. Mod. Phys.  D {\bf 13}, 1321 (2004), astro-ph/0402169;
J. S. Alcaniz and J. A. S. Lima, Phys. Rev. D {\bf 72}, 063516  (2005), astro-ph/0507372;
S. Carneiro and J. A. S. Lima, Int. J. Mod. Phys.  A {\bf 20}, 2465 (2005), gr-qc/0405141;
R. Opher and A. Pelinson, Mon. Not. R. Ast. Soc.  {\bf 362}, 167 (2005);
S. Carneiro, C. Pigozzo and H. A. Borges, Phys. Rev. D {\bf 74}, 023532 (2006).


\bibitem{ML02} J. M. F. Maia and J. A. S. Lima, Phys. Rev. D {\bf 65}, 083513 (2002), arXiv:astro-ph/0112091.

\bibitem{EA1} F. E. M.~Costa, J.~S.~Alcaniz and J.~M.~F.~Maia, Phys.\ Rev.\  D {\bf 77}, 083516 (2008). 

\bibitem{Harko11} S. Basilakos, M. Plionis and J. A. S. Lima, Phys. Rev. D {\bf 82}, 083517 (2010), arXiv:1006.3418; 
S. Basilakos, M. Plionis and  J. Sol\`a,  Phys. Rev. D {\bf 82}  083512  (2010), arXiv:1005.5592;
T. Harko,  F. S. N. Lobo,  Shin'ichi Nojiri and Sergei D. Odintsov, Phys. Rev. D {\bf 84},  024020 (2011).

\bibitem{LBS2012} F. E. M. Costa, J. A. S. Lima and F. A. Oliveira, arXiv:1204.1864v1 [astro-ph.CO];
J. S. Alcaniz, H. A. Borges, S. Carneiro, J. C. Fabris, C. Pigozzo and W. Zimdahl,  Phys. Lett. B {\bf 716},  165 (2012), arXiv:1201.5919;
Pradhan, R. Jaiswal and  R. K. Khare, Astrophys Space Sci. {\bf 343}, 489 (2013);
J. A. S. Lima, S. Basilakos and J. Sol\`a,  Mon. Not. R. Ast. Soc. (2013), In press, arXiv:1209.2802 [gr-qc];  
 

\bibitem{CL12} M. Campos and J. A. S. Lima,  Phy. Rev. D {\bf 86}, 043012 (2012), arXiv:1207.5150 [gr-qc].

\bibitem{naked} K. S. Virbhadra and G. F. R. Ellis, Phys. Rev. D {\bf 65}, 103004 (2002);
Z. Kov\'acs and T. Harko, Phys. Rev. D {\bf 82}, 124047 (2010);
S. Sahu. M. Patil, D. Narashima and P. S. Joshi, Phys. Rev. D {\bf 86}, 063010 (2012). 

\bibitem{AL88} M. J. D. Assad and J. A. S. Lima, Gen. Rel. Grav. {\bf 20}, 527 (1988). 
For a more detailed version see preprint from  Brazilian Center of Research Physics, Brazil - CBPF/NF/050/86 (1986). 

\bibitem{CaiWang} R-G. Cai and A. Wang, Phys. Rev.  D {\bf 73}, 063005 (2006).

\bibitem{VariableG}  O. Bertolami, Nuovo Cimento, {\bf 93} B, 36 (1986);
J. C. Carvalho and J. A. S. Lima, Gen. Rel. Grav. {\bf 26}  909 (1994);
J. Sol\`a and H. Stefancic,  Mod. Phys. Lett. A {\bf 21}, 479 (2006), arXiv:astro-ph/0507110.

\bibitem{SS} I. L. Shapiro and  J. Sol\`a,  JHEP 02, 006 (2002), hep-th/0012227;
J. Phys. A {\bf 41}, 164066 (2008). For a recent review see J. Sol\`a, J. Phys. Conf. Ser. {\bf 283},  012033 (2011), arXiv:1102.1815.
 

\bibitem{Abra}
M. Abramowitz and I. A. Stegun, Handbook of Mathematical Functions (Dover Publications, New York, 1964).

%\bibitem{H} For a finite matter-energy distribution, the choice of the initial condition $H_i = 0$ is possible only in the closed case. In fact, for $b=1$  it follows that $H_i = %-a_i^{-1}\sqrt{1 - k}$.
 

\bibitem{Lima01} J. A. S. Lima, Am. J. Phys. {\bf 69}, 1245 (2001), astro-ph/0109215.

%\bibitem{Penrose}

%\bibitem{Oppenheimer} J. R. Oppenheimer and H. Synyder, Phys. Rev. {\bf 55},
%455 (1939).

%\bibitem{LL} L. Landau and E. M. Lifschitz, {\it The Classical theory of Fields}, Pergammon Press (1985); %

%P. J. E. Peebles, {\it Principles of Cosmology}, Princeton UP (1993).

\bibitem{RP69} R. Penrose, {\it Nuovo Cimento Soc. Ital. Fis.} {\bf 1} , 252, (1969).

\bibitem{Papa}  A. Papapetrou, {\it A Random Walk in Relativity and Cosmology}, edited by N. Dadhich, J. K. Rao, J.V.
Narlikar, and C. V. Vishveshwara (John Wiley \& Sons,
New York, 1985), p. 184.  
%\& Sons, New York, pp 184-191, (1985).
%
\bibitem{Collapse} P. S. Joshi, {\it Global Aspects in Gravitation and Cosmology}, Clarendon, Oxford, (1993);
D. Christodoulou, Ann. Math. {\bf 140}, 607 (1994).
For more recent reviews, see, e.g., R. Penrose, in {\em Black Holes and Relativistic Stars}, edited by R. M. Wald (University of Chicago Press, 1998);
A. Krolak, Prog. Theor. Phys. Suppl. {\bf 136}, 45 (1999);
P. S. Joshi, Pramana {\bf 55}, 529 (2000),
and P.S. Joshi, ``{\em Cosmic Censorship: A Current Perspective}," {\tt gr-qc/0206087} (2002);
{\em Gravitational Collapse End States}, {\tt gr-qc/0412082} (2004), and references therein.
A. Beesham, Pramana {\bf 77}, 429 (2011).
     
%\bibitem{Hawking1} S. W. Hawking, in {\it Black Holes}, edited by C. DeWitt and B. S. DeWitt (Gordon and Breach, New York, 1973).
%
\bibitem{Hawking} S. W. Hawking and G. F. R. Ellis, {\em The Large Scale
Structure of Spacetime}, Cambridge University Press, Cambridge  (1973).

\bibitem{Anninos} P. Anninos {\it et al.} Phys. Rev. D {\bf 50}  3801 (1994).

\bibitem{CM73}  M. E. Cahill and G. C. McVittie, J. Math. Phys. {\bf 11}, 1382 (1970).

\bibitem{McV} G. C. McVittie, Mon.  Not. R. Astron. Soc. {\bf 93}, 325 (1933).

\bibitem{Poisson} E. Poisson and W. Israel, Phys. Rev. D {\bf 41}  1796 (1990);
A. Wang, J. F. Villas da Rocha, and N. O. Santos, Phys. Rev. D  {\bf 56}, 7692 (1997);
J. F. Villas da Rocha, A. Wang, and N. O. Santos, Phys. Lett. A {\bf 255}, 213 (1999);
S. A. Hayward, Phys. Rev. D {\bf 70}, 104027 (2004); Phys. Rev. Lett. {\bf 93}, 251101 (2004). 

%\bibitem{Mortlock}
%D. J. Mortlock {\it et al.},  Nature, {\bf 474}, 616, (2011).


%
%
%
%\bibitem{Malafarina}
%P. S. Joshi and D. Malafarina, Int. J. Mod. Phys. {\bf D20}, 2641 (2011).
%
%
%\bibitem{Virbahadra}
%K. S. Virbahadra and G. F. R. Ellis, Phys. Rev. {\bf D65}, 103004 (2002).
%
%\bibitem{Sahu}
%S. Sahu, M. Patil, D. Narasimha and P. S. Joshi, {\tt arXiv:1206.3077v1} (2012).
%
%\bibitem{Kovac}
%Z. Kov\'acs and T. Harko, Phys. Rev. {\bf D82}, 124047 (2010).
%
%\bibitem{Birkel}
%M. Birkel and S. Sarkar, Astropart. Phys. {\bf 6}, 197 (1997). 
%
%\bibitem{ALima}
%J. A. S. Lima, J. M. F. Maia and N. Pires, IAU Symposium {\bf 198}, 111 (2000).
%
%\bibitem {Lima1996} J. A. S. Lima, Phys. Rev. D {\bf 54}, 2571 (1996),  gr-qc/9605055; 
%{\bf ibdem} Gen. Rel. Grav. {\bf 805}, 27 (1997), gr-qc/9605056;  J. A. S. Lima, A. I. Silva and S. M. Viegas, Mon. Not. R. Astron. Soc. %{\bf 312}, 747
%(2000), astro-ph/9902337.
%
%\bibitem{probes} R. Opher and A. Pelinson, Mon. Not. R. Ast. Soc.  {\bf 362}, 167 (2005); G. Luzzi et al., Astrophys. J. {\bf 705}, 1122 %(2009); P. Noterdaeme at, al. Astron. Astrophys. {\bf 526}, L7 (2011).
%
%\bibitem{Basilakos2}
%S. Basilakos, Astron. and Astrophys. {\bf 508}, 575 (2009).
%
%S. Basilakos and J. A. S. Lima, Phys. Rev. D {\bf 82}, 023504 (2010), arXiv:1003.5754;	
%
\end{thebibliography}
\end{document}